\newcommand{\eqb}{\begin{equation}}
\newcommand{\eqe}{\end{equation}}
\newcommand{\dmb}{\begin{displaymath}}
\newcommand{\dme}{\end{displaymath}}
\newcommand{\pd}{\partial}

\newcommand{\eab}{\begin{eqnarray}}
\newcommand{\eae}{\end{eqnarray}}
\newcommand{\ra}{\right\rangle}
\newcommand{\la}{\left\langle}

\newcommand{\be}{\begin{equation}}
\newcommand{\ee}{\end{equation}}

\newcommand{\La}{\Lambda}
\documentclass{ws-mpla}
\usepackage{bbm}
\begin{document}

\markboth{Ralf Hofmann}
{SU(2) and SU(3) Yang-Mills thermodynamics and some implications}

%%%%%%%%%%%%%%%%%%%%% Publisher's Area please ignore %%%%%%%%%%%%%%
\catchline{}{}{}{}{}
%%%%%%%%%%%%%%%%%%%%%%%%%%%%%%%%%%%%%%%%%%%%%%%%%%%%%%%%%%%%%%%%%%%

\title{SU(2) and SU(3) Yang-Mills thermodynamics and some implications}

\author{Ralf Hofmann}

\address{{\em Institut f\"ur Theoretische Physik\\ 
Universit\"at Heidelberg\\ 
Philosophenweg 16\\ 
69120 Heidelberg, Germany\vspace{0.5cm}\\ 
r.hofmann@thphys.uni-heidelberg.de}}

\maketitle

\pub{Received 30 March 2006}{Revised 24 November 2006}

\begin{abstract}
We sketch the development of effective theories for SU(2) and SU(3) 
Yang-Mills thermodynamics. The most important results are quoted 
and some implications for particle physics and 
cosmology are discussed.

\end{abstract}

\ccode{PACS Nos.: 11.15.Ex; 11.15.Tk; 12.40.Ee}

\section{Introduction}	

There are well-known reasons 
why the thermodynamics of Yang-Mills 
theories should be formulated in a rigorous, 
nonperturbative, and analytical setting: the nonconvergence 
of perturbative loop expansions owing 
(i) to its (at best) asymptotic nature \cite{LargeOrderPT} and 
(ii) to the weak screening of the 
magnetic sector (infrared instability) \cite{Linde1980}. 

In 1975 Polyakov conjectured that the infrared 
instability of Yang-Mills theory can be cured 
by taking into account the spatial correlations provided by 
the topologically nontrivial sector \cite{Polyakov1975}. Notice that 
the weight of a nontrivial, (anti)selfdual configuration in the partition 
function is of the form $\exp\left[-\frac{\mbox{const}}{g^2}\right]$ which 
has an essential singularity at $g=0$: A weak-coupling 
expansion thus ignores these configurations completely.    

The purpose of this paper is to give a brief account of the development and of some 
implications of the effective theories for thermalized SU(2) and 
SU(3) Yang-Mills dynamics. The basic idea is to subject the dynamics of 
topological field configurations to an optimized 
spatial coarse-graining \cite{Hofmann2005}. 
In this way the notion of dynamical ground states 
emerges which break the fundamental gauge symmetry 
in successive stages as temperature decreases. 
There are a deconfining, a preconfining, and a
confining phase for SU(2) and SU(3) 
thermodynamics. The loop-expansion of 
thermodynamical quantities\footnote{This is {\sl not} an expansion in the 
effective gauge 
coupling $e$ but only in $\hbar^{-1}$.} is nontrivial only in 
the deconfining phase. A two-loop calculation for the pressure seems 
to indicate a very rapid numerical convergence \cite{HerbstHofmannRohrer2004,SchwarzHofmannGiacosa2006}. 
Expanding in loops integrates out portions of quantum 
fluctuations which remain after the spatial coarse-graining. 
A fixed-order result predicts what should be regarded a 
quantum fluctuation in the next-order calculation. In the preconfining 
phase the excitations are free but massive dual gauge modes. 
A common characteristic of the deconfining and 
preconfining phases is that their ground-state 
pressures are negative: The dynamics of the ground 
state in each of these two phases 
generates vacuum-energy density which depends on 
temperature $T$ in a linear way. In the confining 
phase the ground-state pressure is precisely zero for 
$T\ll\Lambda$ where $\Lambda$ denotes the 
Yang-Mills scale. For $T\sim\Lambda$ thermal equilibrium 
sq down in the confining phase: The density of 
massive spin-1/2 fermion states (selfintersecting center-vortex loops) 
is over-exponentially rising with energy. 
By necessity this destroys the macroscopic homogeneity of the 
system (vicinity to a Hagedorn transition). 
We believe (but cannot prove at the present stage of development) 
that it is the physics taking place at the onsets of local 
Hagedorn transitions and the fact that the position of an 
intersection point in a center-vortex loop is a 
modulus of this soliton effectively lead to 
quantum mechanical transitions 
in any Standard-Model vertex involving 
charged particles.             

The above-sketched results seem to resolve a number 
of problems both theoretical and empirical in nature. 
On the theoretical side, the infrared instability inherent 
in perturbative loop expansions is cured by an emerging adjoint 
Higgs mechanism after a spatial coarse-graining is performed 
over interacting calorons and anticalorons in the deconfining phase. 
The vanishing of the ground-state pressure in the confining 
phase is relevant for an understanding of the role of (naive) zero-point 
fluctuations in quantum field theories associated with nonabelian 
gauge symmetries. On the empirical side, there is a number of experimentally 
testable predictions arising 
from the postulate SU(2)$_{\tiny\mbox{CMB}}\stackrel{\tiny\mbox{today}}=U(1)_Y$. 
In the case of experimental confirmation of the latter 
the Standard Model's mechanism for electroweak
symmetry breaking is endangered by Big-Bang nucleosynthesis. 

\section{Deconfining phase}

Here we gather the essential steps needed to construct the effective 
theory in the deconfining phase. Explaining some (and not all) technicalities, 
we resort to the case SU(2). Necessary generalizations to SU(3) 
are mentioned in passing. 

To apply a spatial-coarse graining to the highly complex 
dynamics of a nonperturbative ground state at a high 
temperature $T\gg\Lambda$ one first writes a definition 
for the phase $\hat\phi$ of an adjoint scalar field 
$\phi$. In contrast to the (spatially homogeneous) 
modulus $|\phi|$ the phase 
$\hat\phi$ does not carry information on dimensional 
transmutation: in all admissible gauges it is 
periodic in the Euclidean time $\tau$. Therefore the derivation 
of its dynamics must involve exact solutions to the (Euclidean) 
equations of motion of the underlying theory. One can show that $\hat\phi$ is contained in the 
space of functions defined by the right-hand side of the following 
equation \cite{Hofmann2005,HerbstHofmann2004}:
%*********
\eab
\label{defphi}
\hat{\phi}^a\equiv\frac{\phi^a}{|\phi|}(\tau)\sim \int d^3x\,\int d\rho\,\,\frac{\lambda^a}{2}&& 
F_{\mu\nu}[A_\alpha(\rho,\beta)]\left((\tau,0)\right)\,
\left\{(\tau,0),(\tau,\vec{x})\right\}[A_\alpha(\rho,\beta)]\times\nonumber\\ 
&&F_{\mu\nu}[A_\alpha(\rho,\beta)]\left((\tau,\vec{x})\right)\,
\left\{(\tau,\vec{x}),(\tau,0)\right\}[A_\alpha(\rho,\beta)]\,.\nonumber\\ 
\eae
%*********
In Eq.\,(\ref{defphi}) a sum over trivial-holonomy calorons and anticalorons 
(in singular gauge) of topological charge modulus one is implicit, and the 
instanton scale parameter is denoted by $\rho$. The Wilson lines 
$\{(\tau,0),(\tau,\vec{x})\}=\{(\tau,\vec{x}),(\tau,0)\}^\dagger$ are to be performed 
along straight lines. When evaluating the 
right-hand side of Eq.\,(\ref{defphi}) the only relevant term in the integrand is
%********
\eqb
\label{relint}
2\,\frac{x^a}{r}\,\sin(2g(\tau,r))\frac{\pd_r^2\Pi(\tau,r)}{\beta r^3}
\eqe
%*********
where $g\equiv\int_0^1 ds\,\frac{r}{2}\pd_\tau \ln \Pi(t,sr)$, 
$\beta\equiv\frac{1}{T}$, $\Pi$ is the `pre'-potential of the 
(anti)caloron \cite{HS1977}, $r\equiv|\vec{x}|$, and $a=1,2,3$. Since 
only spatial derivatives enter in (\ref{relint}) 
the nontriviality of $\hat\phi$ is associated with 
the magnetic sector of the theory. When evaluating the 
integral in Eq.\,(\ref{defphi}) over the term in (\ref{relint}) ambiguities 
arise. On the one hand, the radial and the integration over $\rho$ diverge. 
On the other hand, the integration over the azimuthal 
angle yields a vanishing result. Regularizing all three 
integrals thus produces an undetermined normalization. 
The important point is that the angular regularization, which involves a particular 
direction $\hat{x}$ in space (breaking of rotational symmetry), is not gauge 
invariant: This would be overly restrictive. Rather, a change 
$\hat{x}\to \hat{x}^\prime$ boils down to a global 
gauge transformation. Since this does not 
affect the physics we conclude that the angular regularization employed 
is admissible. Moreover, one observes that the $\tau$ dependence of the integral in 
Eq.\,(\ref{defphi}) (a sine subject to a phase shift) 
saturates very rapidly when increasing the cutoff for the 
$\rho$-integration to a few times $\beta=1/T$. To summarize, for given 
directions $\hat{x}_C,\hat{x}_A$ (angular regularizations 
for caloron and anticaloron contribution) there are four 
parameters that are undetermined in the integral of Eq.\,(\ref{defphi}): 
two phase shifts $\tau_C$ and $\tau_A$ and two normalizations $C_C$ and $C_A$ 
for the contributions of the caloron and the 
anticaloron, respectively \cite{HerbstHofmann2004}. Each set of ambiguities spans 
the kernel ${\cal K}$ of the linear differential operator 
${\cal D}=\pd_\tau^2+\left(\frac{2\pi}{\beta}\right)^2$: The operator ${\cal D}$ 
thus is uniquely determined. 

To find $\hat\phi$'s equation 
of motion one selects only those members of ${\cal K}$ which are 
BPS saturated, that is, energy- and pressure-free. (A spatial average 
over a noninteracting pair of energy- and pressure-free caloron and anticaloron (owing to 
their (anti)selfduality) needs to be energy- and pressure-free.) 
This is equivalent to $\hat\phi$ satisfying the following first-order equation: 
$\pd_\tau \hat{\phi}=\pm\frac{2\pi i}{\beta}{\hat{\lambda}} \,\hat{\phi}\,$ where 
$\hat{\lambda}$ is a normalized linear combination of the Pauli matrices 
$\lambda_a$. If, for definiteness, we set $\hat{\lambda}=\lambda_3$ 
then the solution to this equation of motion reads $\hat{\phi} = C\, \lambda_1 \, 
\exp\left(\mp \frac{2\pi i}{\beta} \lambda_3 (\tau -\tau_0)\right)$. (The phase 
$\hat{\phi}$ winds along a circle on the group manifold $S_3$.) 
Thus the requirement of BPS saturation reduces the number of undetermined 
parameters from four to two: $C$ and $\tau_0$. 

The modulus $|\phi|$ emerges because a Yang-Mills scale $\Lambda_E$ exists 
on the quantum level. At the same time, the length scale 
$|\phi|^{-1}$ determines the size of the volume over which the spatial 
coarse-graining, involving a caloron and its anticaloron, is performed. 
Still considering the interaction-free situation\footnote{As we will show below this point of view proves to be
selfconsistent.}, a spatial 
coarse-graining down to the resolution $|\phi|$ yields an energy- and pressure-free 
field $\phi$: In the absence of interactions $\phi$ is BPS saturated. Since the 
right-hand side of $\phi$'s BPS equation defines the `square root' of $\phi's$ 
potential $V_E(\phi)$ and since the spatial coarse-graining did not 
invoke an explicit $T$-dependence arising from the 
weight in the truncated partition function (recall, that the classical action of a caloron 
does not depend on $T$) the entire BPS equation must not exhibit an 
explicit $T$-dependence. Away from a phase transition, 
the `square root' of $V_E$, in addition, should be an analytic function 
of $\phi$, and the known $\tau$-dependence of the phase $\hat{\phi}$ must be reproduced. 
Up to global gauge rotations the only viable possibility 
for $\phi$'s BPS equation then is:
%*******
\eqb\label{bps14}
\pd_\tau\phi=\pm i\,\Lambda_E^3\,\lambda_3\,\phi^{-1}\,
\eqe
%**********   
where $\phi^{-1}\equiv \frac{\phi}{|\phi|^2}$. Setting $C=1$ and 
substituting $\phi=|\phi|\hat{\phi}$ into Eq.\,(\ref{bps14}), 
one has $|\phi|(\beta,\Lambda_E)=\sqrt{\frac{\beta\Lambda_E^3}{2\pi}}$: 
The modulus $|\phi|$ falls off with the square root of increasing 
temperature. What is of great importance is that an 
action for the field $\phi$ follows from its 
BPS equation of motion and not vice versa. Namely, `squaring' 
the right-hand side of Eq.\,(\ref{bps14}), one arrives at 
$V_E=\mbox{tr}\frac{\La_E^6}{\phi^2}=4\pi\Lambda_E^3 T$. As a consequence, the usual freedom of 
shifting the potential, $V_E\to V_E+\mbox{const}$, allowed by the 
(second-order) Euler-Lagrange equations, is absent. Another important 
observation is that at a given temperature $T$ the mass squared 
$\pd^2_{|\phi|}\,V_E(|\phi|)$ of possible fluctuations $\delta\phi$ is much 
larger than the resolution squared $|\phi|^2$ and also 
than $T^2$. Thus $\phi$ is an inert background in the effective 
theory \cite{Hofmann2005}: Quantum and statistical fluctuations $\delta\phi$ are absent! 
Both the classical and fluctuation inertness of $V_E(|\phi|)$ will 
lead to a uniquely determined energy density of the ground state.  

After spatial coarse-graining the topologically trivial 
sector $\{a_\mu\}$ of the theory couples in a minimal way to the field 
$\phi$. In the effective action the 
kinetic term for $a_\mu$, $\frac{1}{4}\mbox{tr}\,G^2$ with 
$G_{\mu\nu}=\pd_\mu a_\nu-\pd_\nu a_\mu+ie[a_\mu,a_\nu]$ and $e=e(T)$ being the {\sl effective} 
gauge coupling, is that of the fundamental theory because of the latter's 
renormalizability \cite{'t HooftVeltmann}. Namely, 
when expanded loop by loop in perturbation theory the 
kinetic term $\frac{1}{4}\mbox{tr}\,G^2$ is {\sl form invariant}, that is, the effects of 
integrated-out quantum fluctuations solely reside in the momentum 
dependence of the gauge coupling and of the wave-function 
renormalization. Notice that the momentum that enters 
as an argument of the effective gauge coupling is 
given by the scale $|\phi|=|\phi|(T)$ down to which quantum fluctuations 
have been integrated out. The appearance of a 
wave-function renormalization translates into so-called compositeness 
constraints: if, in a physical gauge and for a given vertex, the momentum transfer is further 
off the mass-shell than the scale $|\phi|$ then 
this vertex does not exist in the effective theory. The issue of 
how strong the allowed interactions are in the effective theory 
was studied in \cite{HerbstHofmannRohrer2004,SchwarzHofmannGiacosa2006}: 
Interactions between (quasiparticle) excitations only contribute 
a fraction $\sim 10^{-3}$ to the total 
thermodynamical pressure (two-loop correction). 
We conjecture here that not much happens to the pressure beyond two-loops in the expansion in 
powers of $\hbar^{-1}$. One the one hand, this is 
suggestive because the compositeness constraints after the 
inclusion of the one-loop on-shell conditions become much tighter 
than on tree-level, probably so tight that at least a large fraction of all naively 
allowed diagrams does not exist beyond two-loop. On the other hand, 
there is an anti-weirdness argument. Beyond two-loops so-called pinch 
singularities take place. That is, one encounters expressions such as 
$\left(\delta(p^2-m^2)\right)^2$ which, mathematically, make no 
sense at all. At least those diagrams that exhibit pinch singularities should be excluded 
by the (dressed) compositeness constraints\footnote{In real-time 
perturbation theory Feynman rules depend on the contour which is employed to analytically 
continue from imaginary time. As a result, $2\times2$-matrix valued 
propagators emerge such that pinch singularities cancel 
one another, see \cite{LandsmanWeert}. This, however, doubles the number of 
degrees of freedom in comparison with the 
imaginary-time formalism! Moreover, it is mathematically not clear 
whether results obtained at a fixed order in the loop expansion 
actually do depend on the choice of the contour 
because certain presuppositions for the 
Riemann-Lebesque lemma, which would guarantee the independence 
of the contour, may not be satisfied.}. In a gauge 
theory the only classical zero-momentum gauge-field 
configuration is pure gauge, $G_{\mu\nu}=0$. 
Indeed, such a configuration $a_\mu^{bg}=\frac{\pi}{e}\,T\,\delta_{\mu 4}\lambda_3$ 
solves the gauge-field equation of 
motion
%*******
\eqb\label{eomts}
D_\mu G_{\mu\nu}=ie[\phi,D_\nu \phi]
\eqe
%*******
of the effective theory. The configuration $a_\mu^{bg}$ represents, in an averaged way, 
all gluon exchanges of momentum transfer larger than $|\phi|$ in 
between calorons and anticalorons and radiative corrections thereof. 
The interesting feature is that $a_\mu^{bg}$ shifts the vanishing 
energy density $\rho^{gs}$ and 
the pressure $P^{gs}$ due to noninteracting 
calorons and anticalorons to finite values in the case of interactions: 
$\rho^{gs}=-P^{gs}=4\pi\Lambda_E^3 T$. This renders the 
so-far hidden scale $\Lambda_E$ (gravitationally) detectable. A 
microscopic interpretation of this result is 
available \cite{Hofmann2005,HerbstHofmann2004} 
owing to the work in \cite{Nahm1984,KraanVanBaalNPB1998,vanBaalKraalPLB1998,LeeLu1998,Diakonov2004}. 

It is useful to investigate the Polyakov 
loop ${\cal P}$ in the effective theory since performing a spatial 
coarse-graining and evaluating ${\cal P}$'s expectation are actions that 
essentially commute\footnote{Notice that this is not true for 
the spatial Wilson loop which measures the spatial string tension.}. 
In the gauge, where the field $\phi$
winds around the group manifold $S_3$, one has: 
${\cal P}[a_\mu^{bg}=\frac{\pi}{e}\,T \delta_{\mu 4}\,\lambda_3]=-{\bf 1}$. 
In unitary gauge, which is reached by an admissible albeit 
singular gauge transformation \cite{Hofmann2005}, one obtains: 
${\cal P}[a_\mu^{bg}=0]={\bf 1}$. As a consequence, the 
ground state is degenerate w.r.t. an electric $Z_2$ 
symmetry. This statement remains valid on the level of the 
total expectation $\la {\cal P}\ra$ because the field $\phi$ breaks the 
gauge symmetry as SU(2)$\to$U(1) thus saving the masslessness of one color direction. 
We conclude that the discussed phase is, indeed, 
deconfining. For SU(3) one obtains an adjoint scalar field $\phi$ which winds in 
each of the (not entirely independent) SU(2) subalgebras for a third of the 
time $0\le\tau\le\beta$. The Polyakov-loop expectation now is degenerate w.r.t. a 
$Z_3$ electric symmetry, and two directions in color space 
remain massless. For SU(N) with $N\ge 4$ no unique phase diagram 
exists with the possible exception\footnote{We believe that the 
gauge symmetry SU($\infty$) is a progenitor for diffeomorphism 
invariance emerging in the confining phase.} of $N=\infty$.  

Let us now briefly describe the physics of excitations. By the adjoint 
Higgs mechanism and in unitary gauge $\phi=\mbox{const}, a^{bg}_\mu=0$ two (six) out 
of three (eight) color direction acquire a mass for SU(2) (SU(3)): 
Gluons become thermal quasiparticle excitations on tree-level 
by coarse-grained interactions with the ground state. For SU(2) 
one has $m=2e|\phi|$, for SU(3) $m_2=2e|\phi|$ and $m_4=e|\phi|$. 
Massive gauge bosons provide for infrared cutoffs and thus 
cure the old perturbative instability problem of the magnetic sector. To learn 
how the effective gauge coupling $e$ depends on the cutoff $|\phi|$ 
(and thus on temperature $T$) one imposes the requirement that the pressure and energy density 
are related by one and the same Legendre transformation, regardless of whether one 
chooses to calculate them in fundamental or effective 
field variables. On the level of free quasiparticles, 
which is sufficient for many practical situations, one 
arrives at the following evolution equation: $\pd_a \lambda_E=-\frac{24\lambda_E^4 a}{(2\pi)^6}\frac{D(2a)}
{1+\frac{24\lambda_E^3 a^2}{(2\pi)^6}}$
where $\lambda_E\equiv\frac{2\pi T}{\Lambda_E}$, $a\equiv\frac{m}{2T}$, and 
$D(a)\equiv\int_0^\infty dx\,\frac{x^2}{\sqrt{x^2+a^2}}\,\frac{1}{\exp(\sqrt{x^2+a^2})-1}\,$. 
Notice that $e$, $\lambda$, and $a$ are related as: 
$a=2\pi\,e\,\lambda_E^{-3/2}\,$. As a result of the one-loop evolution, 
$e$ is essentially constant for $\lambda_E$ sufficiently larger than 
$\lambda_{c,E}$. One has $e=8.89,\,\lambda_{c,E}=13.87$ (SU(2)) 
and $e=7.26,\,\lambda_{c,E}=9.475$ (SU(3)). 
For $\lambda_E\searrow\lambda_{c,E}$ the coupling diverges in a
logarithmic fashion: $e\sim-\log(\lambda_E-\lambda_{c,E})$. As a consequence, stable and screened 
magnetic monopoles, which are isolated objects deep inside the 
preconfining phase\cite{Hofmann2005,Diakonov2004}, become massless at 
$\lambda_{c,E}$ and thus condense. The associated phase transition 
is second-order like \cite{Hofmann2005}. It is worth 
mentioning that the entire thermodynamics of the deconfining phase 
is robust against changes of initial conditions set at a 
sufficiently high temperature: A decoupling of ultraviolet from infrared 
physics takes place. To set $m=0$ for $T>T_i\gg T_{c,E}$, where $m(T_i)=0$ is the initial 
condition for the (downward) evolution of the effective coupling $e$, and to still believe 
in the physical relevance of four-dimensional Yang-Mills gauge-field theory is a contradiction: 
As we have learned, even at a high temperature this dynamics is only consistent 
nonperturbatively. We thus conclude that $T_i$ must 
coincide with the temperature where the concept of a 
four-dimensional, smooth spacetime manifold breaks down. 
There are good reasons to believe that $T_i\sim 1.2\times 10^{19}\,$GeV. 

\section{Preconfining phase}

Let us know discuss what happens at and slightly below $T_{c,E}$. Because of total screening, 
magnetic monopoles (one species for SU(2), two species for
SU(3)) become massless, and thus are extremely abundant: All memory of the existence of the 
Yang-Mills scale $\Lambda$ is washed away at $T_{c,E}$. The theoretical challenge is to 
describe the associated, newly emerging ground-state physics in 
an analytical way. It is clear that due to the highly complex dynamical situation 
exact results, again, require the process of a spatial coarse-graining.  

The situation in the case of SU(2) is the following: 
The key is to consider the mean magnetic flux $\bar{F}_{\pm,\tiny\mbox{thermal}}$ of a noninteracting 
monopole-antimonopole pair of vanishing 
spatial momentum (also for each constituent) 
through an $S_2$ of infinite radius in a thermal environment. (A finite radius would be associated 
with a mass scale which does not yet exist.) One has
%*****
\eab
\label{avfluxsys}
\bar{F}_{\pm,\tiny\mbox{thermal}}(\delta)&=&
\,4\pi\,\int d^3p\,\delta^{(3)}(\vec{p})\, n_B(\beta |E(\vec{p})|)\,\bar{F}_{\pm}\nonumber\\ 
&=&\pm\frac{8\pi\,\delta}{e}\int d^3p\,
\frac{\delta^{(3)}(\vec{p})}{\exp\left[\beta\sqrt{M_{m+a}^2+\vec{p}^2}\right]-1}\,
\eae
%*********
where $\bar{F}_{\pm}\equiv\pm\frac{\delta}{2\pi}\frac{4\pi}{e}=
\pm\frac{2\delta}{e}$, and $\delta\,,\ \ (0\le\delta\le\pi),$ is the 
angle between the Dirac strings of 
the monopole and antimonopole. (We work in unitary gauge on the microscopic level.) After setting $\vec{p}=0$ 
(spatial average) in $\left(\exp\left[\beta\sqrt{M_{a+b}^2+\vec{p}^2}\right]-1\right)$ and 
with $M_{a+b}=\frac{8\pi^2}{e\beta}$ (after screening
\cite{KraanVanBaalNPB1998,vanBaalKraalPLB1998,LeeLu1998}), 
the expansion of this term reads
%*********
\eqb
\label{expexpon}
\lim_{\vec{p}\to 0}\left(\exp\left[\beta\sqrt{M_{m+a}^2+\vec{p}^2}\right]-1\right)=
\frac{8\pi^2}{e}\left(1+\frac{1}{2}\frac{8\pi^2}{e}+
\frac{1}{6}\left(\frac{8\pi^2}{e}\right)^2+\cdots\right)\,.
\eqe
%********
The limit $e\to\infty$ can safely be performed in Eq.\,(\ref{avfluxsys}), 
and we have
%**********
\eqb
\label{avfluxsys,einf}
\lim_{e\to\infty} \bar{F}_{\pm,\tiny\mbox{thermal}}(\delta)=
\pm\frac{\delta}{\pi}\,,\ \ \ \ \ \ (0\le\delta\le\pi)\,.
\eqe
%********** 
This is finite and depends on the 
angular variable $\delta$ continuously. 
Now $\frac{\delta}{\pi}$ is a (normalized) angular variable just 
like $\frac{\tau}{\beta}$ is. 
Thus we may set $\frac{\delta}{\pi}=\frac{\tau}{\beta}$. 
The macroscopic complex field $\varphi$, describing the monopole 
condensate in the absence of interactions, is spatially 
homogeneous  and 
its phase $\hat\varphi\equiv\frac{\varphi}{|\varphi|}$ 
depends on $\frac{\tau}{\beta}$ only and in a periodic way. 
Moreover, 
since the physical flux situation for the thermalized monopole-antimonopole pair 
does not repeat itself for $0\le \frac{\delta}{\pi}\le 1$ we conclude that 
this period is $\pm$ unity: 
%*********
\eqb
\label{varphiphase}
\hat\varphi(\tau)=C\exp\left[\pm 2\pi i\frac{\tau-\tau_0}{\beta}\right]\, 
\eqe
%**********
where $C$ and $\tau_0$ are undetermined. 

To derive $\varphi$'s modulus, which together with $T$ determines the length scale 
$|\varphi|^{-1}$ over which the spatial average is performed, 
we proceed in close analogy to the deconfining phase. That is, 
we assume the existence of an (at this stage) externally 
provided mass scale $\Lambda_M$. Since the weight for integrating 
out massless and noninteracting monopole-antimonopole systems in the partition 
function is $T$ independent and since the cutoff in length for the spatial 
average defining $|\varphi|$ is $|\varphi|^{-1}$, an explicit $T$ 
dependence ought not arise in any quantity being 
derived from such a coarse-graining. That is, in the effective action density any $T$ 
dependence (still assuming the absence of interactions between massless monopoles and 
antimonopoles when performing the coarse-graining) must appear through 
$\varphi$ only. Moreover, integrating massless and momentum-free monopoles 
and antimonopoles into the field $\varphi$ means that this field is energy- 
and pressure-free: $\varphi$'s $\tau$ dependence (residing in its phase) 
must be BPS saturated. On the right-hand 
side of $\varphi$'s or $\bar{\varphi}$'s ($\varphi$'s complex conjugate) 
BPS equation the requirement of analyticity (because away from a phase 
transition the monopole condensate should exhibit a smooth $T$ dependence) 
and linearity in $\varphi$ or $\bar{\varphi}$ (because the $\tau$ dependence of $\varphi$'s 
phase, see Eq.\,(\ref{varphiphase}), needs to honoured) yields 
the following first-order equation of motion
%*******
\eqb
\label{BPSvarphi}
\pd_\tau \varphi=\pm i\frac{\La_M^3\,\varphi}{|\varphi|^2}=\pm i\frac{\La_M^3}{\bar{\varphi}}\,.
\eqe
%*******
Substituting $\varphi=|\varphi|\hat\varphi$ into Eq.\,(\ref{BPSvarphi})  
and appealing to Eq.\,(\ref{varphiphase}) (setting $C=1$), we derive $|\varphi|=\sqrt{\frac{\La_M^3\beta}{2\pi}}$. 
Notice that the `square' of the right-hand side in Eq.\,(\ref{BPSvarphi}) uniquely 
defines $\varphi$'s potential $V_M$. (In contrast to a second-order equation of 
motion, following from an action by means of the variational 
principle, Eq.\,(\ref{BPSvarphi}) does {\sl not} 
allow for a shift $V_M\to V_M+\mbox{const}$.) For the case SU(3) the BPS equation 
(\ref{BPSvarphi}) emerges for each of the two independent monopole 
condensates $\varphi_1$ and $\varphi_2$. 

Again, by comparing the curvature of their 
potentials with the square of temperature and the squares of 
their moduli, one concludes that that the field $\varphi$ (SU(2)) and the fields $\varphi_1$, 
$\varphi_2$ (SU(3)) neither fluctuate on-shell nor off-shell\cite{Hofmann2005}: Spatial coarse-graining 
over nonfluctuating, classical configurations generates inert 
macroscopic fields.

To derive the full effective theory the spatially coarse-grained and topologically 
trivial (dual) gauge fields $a^D_\mu$ (SU(2)) and $a^D_{\mu,1}$, $a^D_{\mu,2}$ (SU(3)) are 
minimally coupled (with a universal {\sl effective} magnetic coupling $g$) 
to the inert fields $\varphi$ (SU(2)) and 
$\varphi_1$, $\varphi_2$ (SU(3)). The kinetic terms for $a^D_\mu$ (SU(2)) 
and $a^D_{\mu,1}$, $a^D_{\mu,2}$ (SU(3)) are canonical\footnote{A spatial coarse-graining over {\sl free} 
plane waves does not alter their kinetic term.}. Since the effective theory is abelian 
with (spontaneously broken) gauge group U(1)$_D$ (SU(2)) and U(1)$^2_D$ (SU(3)) 
and since the monopole fields do not fluctuate it follows that 
thermodynamical quantities are exact on the 
one-loop level. Before we discuss the 
spectrum of quasiparticles running in the loop 
we need to derive the full ground-state 
dynamics in the effective theory. The classical equations of motion for 
the dual gauge field $a^D_\mu$ are
%*********
\eqb
\label{eomdualG2}
\pd_\mu G^D_{\mu\nu}=ig\left[\overline{{\cal D}_{\nu}\varphi}\varphi-\bar{\varphi}
{\cal D}_{\nu}\varphi\right]
\eqe
%*********
where $G^D_{\mu\nu}=\pd_\mu a^D_\nu-\pd_\nu a^D_\mu$ and 
${\cal D}_{\mu}\equiv \pd_\mu+ig\,a^D_\mu$. (For SU(3) the 
right-hand sides for the two equations for the dual gauge fields $a^D_{\mu,1}$, $a^D_{\mu,2}$ can be 
obtained by the substitutions $\varphi\to\varphi_1$ or 
$\varphi\to\varphi_2$ in Eq.\,\ref{eomdualG2}.) The pure-gauge solution 
to Eq.\,(\ref{eomdualG2}) with ${\cal D}_\mu\varphi\equiv0$ is given as 
$a^{D,bg}_{\mu}=\pm\delta_{\mu 4}\frac{2\pi}{g\beta}$. 
In analogy to the deconfining phase, the coarse-grained 
manifestation $a^{D,bg}_{\mu}$ of monopole-antimono\-pole in\-teractions, 
mediated by dual, off-shell plane-wave modes on 
the microscopic level, shifts the energy density $\rho^{gs}$ and the 
pressure $P^{gs}$ of the ground state from zero to finite values: 
$\rho^{gs}=-P^{gs}=\pi\,\La_M^3\,T$ (SU(2)) and 
$\rho^{gs}=-P^{gs}=2\pi\,\La_M^3\,T$ (SU(3)). 

In contrast to the 
deconfining phase, where $P^{gs}<0$ arises 
from monopole-antimonopole attraction, the negative ground-state pressure 
in the preconfining phase originates from collapsing and re-created 
center-vortex loops \cite{Hofmann2005}. (There are 
two species of such loops for SU(3) and one species for SU(2)). The core of 
a given center-vortex loop can be pictured as a stream of the associated monopole 
species flowing oppositely directed to the stream of their 
antimonopoles \cite{Olejnik1997}. Since by Stoke's theorem the magnetic flux carried 
by the vortex is determined by the dual 
gauge field $a^{D,\tiny\mbox{tr}}_\mu$ transverse to the vortex-tangential 
and since $a^{D,\tiny\mbox{tr}}_\mu$ is -- in a covariant gauge -- 
invariant under collective boosts of the streaming 
monopoles or antimonopoles in the vortex core it 
follows that the magnetic flux solely depends 
on the monopole charge and not on the collective state of 
monopole-antimonopole motion. This, in turn, implies a 
center-element classification of the magnetic fluxes carried 
by the vortices justifying the name center-vortex loop. Viewed on the 
level of large-holonomy calorons an unstable center-vortex loop is created within a 
region where the mean axis for the dissociation of several calorons represents a net 
direction for the monopole-antimonopole flow. Notice that each so-generated vortex 
core must form a closed loop due to the absence of isolated magnetic 
charges within the monopole condensate. In contrast to the deconfining phase a 
rotation to (macroscopic) unitary gauge $a^{D,bg}_{\mu}=0, \varphi=|\varphi|$ is facilitated 
by a smooth, periodic gauge transformation which leaves the value ${\bf 1}$ 
of the Polyakov loop invariant: The electric $Z_2$ degeneracy, observed 
in the deconfining phase, is lifted in the ground-state. (For SU(3) it 
is an electric $Z_3$ degeneracy that is lifted.) By the dual (abelian) Higgs 
mechanism the mass of (noninteracting) 
quasiparticle modes is given as: $m=g|\varphi|$ (SU(2)) and 
$m_1=g|\varphi_1|=g|\varphi_2|=m_2$ (SU(3)). 

The evolution of the effective magnetic coupling $g$ is determined (for both SU(2) and SU(3)) 
by the equation $\pd_a \lambda_M=-\frac{12\lambda_M^4 a}{(2\pi)^6}\frac{D(a)}{1+\frac{12\lambda_M^3 a^2}{(2\pi)^6}D(a)}$ 
where $\lambda_M\equiv\frac{2\pi T}{\Lambda_M}$, $a\equiv\frac{m}{T}$, 
and $a=2\pi g\lambda_M^{-3/2}$. Continuity of the 
pressure at $T_{c,E}$ relates the scales $\Lambda_M$ and $\Lambda_E$ 
as: $\Lambda_E\sim (1/4)^{1/3}\Lambda_M$ for SU(2) and 
$\Lambda_E\sim (1/2)^{1/3}\Lambda_M$ for SU(3). 
The coupling $g$ rapidly rises from zero at 
$\lambda_{c,E}$ (corresponding to $\lambda_M=8.478$ (SU(2)) and 
$\lambda_M=7.376$ (SU(3))) to infinity at $\lambda_{c,M}=7.075$ (SU(2)) 
and at $\lambda_{c,M}=6.467$ (SU(3)): 
$g\sim -\log(\lambda_M-\lambda_{c,M})$. We conclude that the preconfining 
phase occupies only a narrow region in the phase diagram of each theory. 
At $\lambda_{c,M}$ the core of a center 
vortex loop exhibits a vanishing diameter, and 
the pressure outside of the core vanishes: Single center-vortex 
loops become massless and stable 
spin-1/2 excitations. 

\section{Confining phase}

Here were are interested in the physics taking place 
below $T_{c,M}$. At $T_{c,M}$ single center-vortex loops are extremely 
abundant: They conspire to form a newly 
emerging ground state subject to complex internal dynamics. 
To derive the dynamics of a macroscopic, 
complex field $\Phi$ describing this situation, again, a spatial 
coarse-graining is needed. 

In the absence of interactions between center-vortex loops 
(only contact interactions are possible due to the decoupling 
of the dual gauge modes at $T_{c,M}$) the phase of this field is determined by the 
quantum statistical average flux through an $S_1$ of 
infinite radius, centered at the spatial point $\vec{x}$. 

At $T_{c,M}$ the average flux due to a system $S$ 
of a center-vortex loop and its flux-reversed partner, both at rest, 
reads
%*****
\eab
\label{avfluxsysV}
\lim_{g\to\infty}\,F_{\tiny\mbox{thermal}}&=&4\pi F\,\int d^3p\,\,\delta^{(3)}(\vec{p})\, n_B(\beta |2\,E_v(\vec{p})|)\nonumber\\ 
&=&0,\pm\frac{8\pi}{\beta|\varphi|}=0,\pm 4\,\lambda^{3/2}_{c,M}\,
\eae
%********* 
where $F$ is (the vanishing) flux of the system when not coupled 
to the heat bath. 
According to Eq.\,(\ref{avfluxsysV}) there are finite, discrete, and dimensionless 
parameter values for the description of the 
macroscopic phase 
%********
\eqb
\label{phasePhi}
\Gamma\frac{\Phi}{|\Phi|}(\vec{x})\equiv \lim_{g\to\infty}\la \exp[i\oint_{C(\vec{x})}dz_\mu\, 
a^D_\mu]\ra\ \ \ \ \ (\mbox{'t Hooft loop})
\eqe
%**********
associated with the Bose condensate of the system $S$. 
In Eq.\,(\ref{phasePhi}) $\Gamma$ is an undetermined and dimensionless 
complex constant and $C(\vec{x})$ is the contour described by an $S_1$ of infinite radius. 
For convenience we normalize the parameter 
values arising in $\lim_{g\to\infty}\,F_{\tiny\mbox{thermal}}$ (Eq.\,(\ref{avfluxsysV})) 
as $\hat{\tau}\equiv 0,\pm 1$. 

To investigate the decay of the monopole condensate at $T_{c,M}$ (pre- and reheating) 
and the subsequently emerging equilibrium situation, we need to find conditions to constrain 
the potential $V_C$ for the macroscopic field $\Phi$ in such a way 
that the dynamics arising from it is unique. 
The entire (fermionic) pre- and reheating in the confining phase 
is described by spatially and temporally discontinuous changes of the 
modulus (energy loss) and phase (flux creation) of 
the field $\Phi$. Since the condensation of 
the system $S$ renders the expectation of the 't Hooft loop finite (proportional 
to $\Phi$) the magnetic center symmetry $Z_2$ (SU(2)) and $Z_3$ (SU(3)) 
is dynamically broken as a discrete gauge symmetry. Thus, 
after return to equilibrium, the ground state of the confining phase 
must exhibit $Z_2$ (SU(2)) and $Z_3$ (SU(3)) degeneracy. This implies that for SU(2) 
the two parameter values $\hat{\tau}=\pm 1$ need to be identified while each of the 
three values $\hat{\tau}=\pm 1,0$ describe a distinct 
ground state for SU(3). 

Let us now discuss how either one of 
these degenerate ground states is reached. Spin-1/2 particle creation 
proceeds by single center vortex loops being sucked-in from 
infinity. (The overall pressure is still negative during the 
decay of the monopole condensate thus facilitating the in-flow 
of spin-1/2 particles from spatial infinity.) At a given point $\vec{x}$ an observer 
detects the in-flow of a massless fermion in terms of the field 
$\Phi(\vec{x})$ rapidly changing its phase by a forward center jump 
(center-vortex loop gets pierced by $C(\vec{x})$) which is followed by the associated 
backward center jump (center-vortex loop lies inside $C(\vec{x})$). 
Each phase change corresponds to a tunneling 
transition in between regions of positive curvature in 
$V_C$. If a phase jump has 
taken place such that the subsequent potential 
energy for the field $\Phi$ is still positive then 
$\Phi$'s phase needs to perform additional jumps in order to 
shake off $\Phi$'s energy completely. This can only 
happen if no local minimum exists at a finite value of $V_C$. 
If the created single center-vortex loop moves sufficiently fast it can subsequently 
convert some of its kinetic energy into mass by twisting: massive, selfintersecting 
center-vortex loops arise. These particles are also spin-1/2 
fermions: A $Z_2$ or $Z_3$ monopole, constituting the 
intersection point, reverses\footnote{A particle with $n$ 
selfintersections of the center-vortex loops corresponds 
to one of the distinct topologies in 
the connected vacuum diagrams of a 
$\lambda\phi^4$-theory. One can draw a continuous and closed line along 
the center-flux running around the diagram.} the center 
flux \cite{Reinhardt2001}.
If the SU(2) (or SU(3)) pure 
gauge theory does not mix with any other preconfining or deconfining 
gauge theory, whose propagating gauge modes would couple 
to the $Z_2$ (or $Z_3$) charges, a soliton generated 
by $n$-fold twisting is stable in isolation and possesses a 
mass $n\,\Lambda_C$. Here $\Lambda_C$ is the mass of the 
charge-one state (one selfintersection). After a sufficiently large and even number 
of center jumps has occurred the 
field $\Phi(\vec{x})$ settles in one of 
its minima of zero energy density. 

Let us summarize the results of our above discussion: (i) the potential 
$V_C$ must be invariant under magnetic 
center jumps $\Phi\rightarrow\exp[\pi i]\Phi$\,\,(SU(2)) and 
$\Phi\rightarrow\exp[\pm \frac{2\pi i}{3}]\Phi$\,\,(SU(3)) 
only. (An invariance under a larger continuous or 
discontinuous symmetry is excluded.) (ii) spin-1/2 fermions are created by a forward and 
a backward tunneling corresponding to local center jumps in $\Phi$'s 
phase. (iii) The minima of 
$V_C$ need to be at zero-energy density and are 
all related by center transformations, 
no additional minima exist. (iv) 
Moreover, we insist on the occurrence of one mass 
scale $\Lambda_C$ only to parameterize the potential $V_C$. (As it was 
the case for the ground-state physics in the de - 
and preconfining phases.) 
(v) In addition, it is clear that the potential $V_C$ needs to 
be real.\vspace{0.1cm}\\  
\noindent\underline{SU(2) case:}\vspace{0.1cm}\\ 
A generic potential $V_C$ satisfying (i),(ii), (iii), (iv), and (v) is given by
%*********
\eqb
\label{2potC}
V_C=\overline{v_C}\,v_C\equiv\overline{\left(\frac{\Lambda_C^3}{\Phi}-\La_C\,\Phi\right)}\,
\left(\frac{\Lambda_C^3}{\Phi}-\La_C\,\Phi\right)\,.
\eqe
%*********
The zero-energy minima of $V_C$ 
are at $\Phi=\pm \Lambda_C$. It is clear that 
adding or subtracting powers $(\Phi^{-1})^{2l+1}$ or 
$\Phi^{2k+1}$ in $v_C$, where $k,l=1,2,3,\cdots$, 
generates additional zero-energy 
minima some of which are {\sl not} related by center transformations 
(violation of requirement (iii)). Adding $\Delta V_C$, defined as 
an {\sl even}  power of a Laurent expansion 
in $\bar{\Phi}\Phi$, to $V_C$ (requirements (iii) and (v)), does in general 
destroy property (iii). A possible exception is 
%*******
\eqb
\label{DeltaVC}
\Delta V_C=\lambda\left(\Lambda_C^2-
\Lambda_C^{-2(n-1)}\left(\bar{\Phi}\Phi\right)^n\right)^{2k}
\eqe
%*********
where $\lambda>0; k=1,2,3,\cdots; n\in {\bf Z}$. Such a term, however, 
is irrelevant for the description of the tunneling 
processes (requirement (ii)) since the associated euclidean 
trajectories are essentially along U(1) Goldstone 
directions for $\Delta V_C$ due to the 
pole in Eq.\,(\ref{2potC}). Thus adding $\Delta V_C$ does not cost 
much additional euclidean action and therefore does not affect 
the tunneling amplitude in a significant way. As for the 
curvature of the potential at its minima, adding $\Delta V_C$ does not lower 
the value as obtained for $V_C$ alone. One 
may think of multiplying $V_C$ with a positive, dimensionless 
polynomial in $\Lambda_C^{-2}\bar{\Phi}\Phi$ with 
coefficients of order unity. This, however, 
does not alter the physics of the pre - and reheating 
process. It increases the curvature of the 
potential at its zeros and therefore does not alter the result that quantum 
fluctuations $\delta\Phi$ are absent after relaxation, see below.    
\vspace{0.1cm}\\   
\noindent\underline{SU(3) case:}\vspace{0.1cm}\\ 
A generic potential $V_C$ satisfying (i),(ii), (iii), (iv), and (v) is given by
%*********
\eqb
\label{3potC}
V_C=\overline{v_C}\,v_C\equiv\overline{\left(\frac{\Lambda_C^3}{\Phi}-\Phi^2\right)}\,
\left(\frac{\Lambda_C^3}{\Phi}-\Phi^2\right)\,.
\eqe
%*********
The zero-energy minima of $V_C$ 
are at $\Phi=\Lambda_C\exp\left[\pm\frac{2\pi i}{3}\right]$ and $\Phi=\Lambda_C$. 
Again, adding or subtracting powers $({\Phi}^{-1})^{3l+1}$ or $(\Phi)^{3k-1}$ 
in $v_C$, where $l=1,2,3,\cdots$ and $k=2,3,4,\cdots$, 
violates requirement (iii). The same discussion for adding 
$\Delta V_C$ to $V_C$ and for 
multiplicatively modifying $V_C$ applies as in the SU(2) case. 

\begin{figure}
\begin{center}
\leavevmode
\leavevmode
%\epsffile[80 25 534 344]{}
\vspace{5.5cm}
\includegraphics{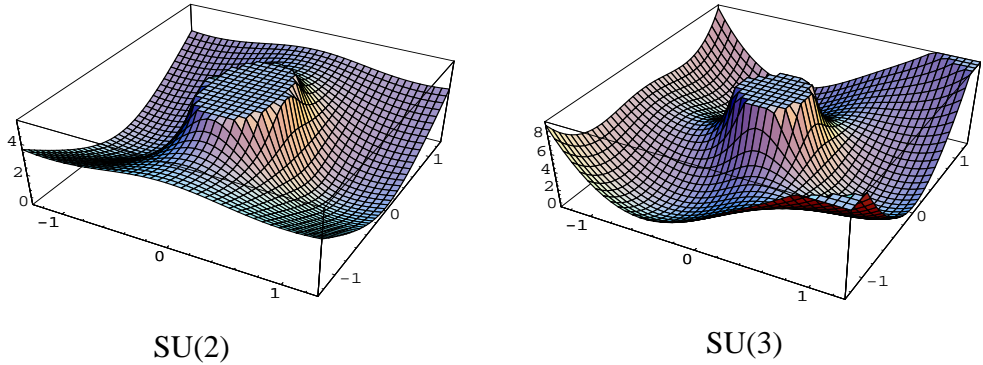}
\end{center}
\caption{The potential $V_C=\overline{v_C(\Phi)}v_C(\Phi)$ for the center-vortex 
condensate $\Phi$. Notice the regions of negative tangential curvature 
in between the minima.\label{Fig-1}}      
\end{figure}

In Fig.\,\ref{Fig-1} plots of the potentials in Eq.\,(\ref{2potC}) 
and Eq.\,(\ref{3potC}) are shown. 
%***********************
%************************
The ridges of negative tangential curvature are classically forbidden: 
The field $\Phi$ tunnels through these ridges, and a 
phase change, which is determined by an element of the center $Z_2$ (SU(2)) or $Z_3$ (SU(3)), 
occurs locally in space. This is the afore-mentioned generation of 
one unit of center flux.

To decide about the absence of fluctuations in each of the 
minima of the potential $V_C$ the following consideration applies. First of all, 
one compares the curvature at the minima with the square of the 
coarse-graining scale $|\Phi_{\tiny\mbox{min}}|=\Lambda_C$. 
Setting $\Phi=|\Phi|\,\exp(i\frac{\theta}{|\Phi|})$ one has
%*******
\eqb
\label{minimacur}
\left.\frac{\pd^2_{\theta} V_C(\Phi)}{|\Phi|^2}\right|_{\Phi_{\tiny\mbox{min}}}=
\left.\frac{\pd^2_{|\Phi|} V_C(\Phi)}
{|\Phi|^2}\right|_{\Phi_{\tiny\mbox{min}}}
=\left\{\begin{array}{c}8\,\ \ \ \ \ (\mbox{SU(2)})\\ 
18\,\ \ \ \ \ (\mbox{SU(3)})\end{array}\right.\,.
\eqe
%*******
Thus a potential fluctuation $\delta\Phi$ would 
be harder than the maximal resolution $|\Phi_{\tiny\mbox{min}}|$ corresponding to 
the effective action that arises after 
spatial coarse-graining, and therefore $\delta\Phi$ is already 
contained in the classical configuration 
$\Phi_{\tiny\mbox{min}}$: $\delta\Phi$ does not exist in the effective theory. Second, we need to 
investigate whether tunneling events take place that lift the 
degeneracy of the minima. According to the WKB-method the probability of a 
tunneling event to another minimum is given by $\exp(-S_{\tiny\mbox{tunn}})$ 
where the subscript `tunn' refers to a classical, 
euclidean trajectory connecting the minima. This trajectory must be 
BPS saturated since the tunneling event itself is 
spontaneous: no energy is needed (and apriori present) for it to occur. 
In \cite{Hofmann2000} BPS saturated trajectories 
subject to the potentials in Eqs.\,(\ref{2potC}) 
and (\ref{3potC}) have been investigated. The only solution to the 
BPS equation, subject to the initial 
condition $\Phi=\Phi_{\tiny\mbox{min}}$, is 
$\Phi\equiv\Phi_{\tiny\mbox{min}}$ for any finite amount of euclidean time $\tau$. 
Since there are no fluctuations 
$\delta\Phi$ available that could possibly alter the 
initial condition $\Phi=\Phi_{\tiny\mbox{min}}$ we conclude 
that tunneling processes are entirely absent. (Viewed alternatively, one would have to wait 
infinitely long for a tunneling process to occur.) We conclude that once 
the field $\Phi$ has relaxed to either one of the minima 
of the potential $V_C$ no quantum fluctuations are around that 
could possibly shift the vanishing ground-state pressure and 
energy density of the theory. Needless to say, this 
result is relevant for an understanding of the 
smallness of today's value of the cosmological 
constant on particle-physics scales. 

The density of massive spin-1/2 states, that is generated in the 
process of relaxation, is over-exponentially increasing. Namely, 
the multiplicity of massive fermion states, associated with center-vortex loops possessing 
$n$ selfintersections, is given by twice the number $L_n$
of bubble diagrams with $n$ vertices in a scalar $\lambda \phi^4$ theory. 
(In the absence of propagating gauge
modes, possibly provided by another Yang-Mills theory, we disregard charge multiplicities.) 
In \cite{BenderWu1969} the minimal number of such diagrams $L_{n,min}$ was 
estimated to be 
%******
\eqb
\label{NOD}
L_{n,min}=n!3^{-n}\,.
\eqe
%******
The mass spectrum is equidistant: The mass $m_n$ of a state with $n$ selfintersections of 
the center-vortex loop is $m_n\sim n\,\La_C$. If we only 
ask for an estimate of the density of {\sl static} 
fermion states $\rho_{n,0}=\tilde{\rho}(E=n\La_C)$ of energy 
$E$ then, by appealing to Eq.\,(\ref{NOD}) 
and Stirling's formula, 
we obtain \cite{Hofmann2005}
%******
\eab
\label{statdes}
\rho_{n,0}&>&\frac{\sqrt{8\pi}}{3\La_C}\,\exp[n\log n]\Big(\log n+1\Big)\,\ \ \ \ \ \mbox{or}\nonumber\\ 
\tilde{\rho}(E)&>&\frac{\sqrt{8\pi}}{3\La_C}\exp[\frac{E}{\La_C}\log\frac{E}{\La_C}]
\Big(\log\frac{E}{\La_C}+1\Big)\,.
\eae
%*******  
The partition function 
$Z_{\Phi}$ for the system of static fermions thus is estimated as
%******
\eab
\label{partfunctionphi}
Z_{\Phi}&>&\int_{E^*}^\infty dE\,\tilde{\rho}(E)\,n_F(\beta E)\nonumber\\ 
&>&\frac{\sqrt{8\pi}}{3\La_C}\,\int_{E^*}^\infty dE\,\exp\left[\frac{E}{\La_C}\right]\,
\exp[-\beta E]\,,
\eae
%*******
where $E^*\gg \Lambda_C$ is the energy where we start to trust 
our approximations. Thus $Z_{\Phi}$ diverges at some 
temperature $T_H<\Lambda_C$. Due to the logarithmic factor in the exponent 
arising in estimate Eq.\,(\ref{statdes}) for $\tilde{\rho}(E)$ we 
would naively conclude that $T_H=0$. This, however, is an artefact 
of our assumption that all states with $n$ selfintersections 
are absolutely stable. Due to the existence of contact interactions between 
vortex lines and intersection points this assumption is 
the less reliable the higher the total energy of a 
given fluctuation. (A fluctuation of large energy has a higher 
density of intersection points and vortex lines and 
thus a larger likelihood for the occurrence of contact interactions which mediate the 
decay or the recombination of a given state with $n$ selfintersections.) 
At the temperature $T_H$ the entropy 
wins over the Boltzmann suppression in energy, and the partition 
function diverges. To reach the point $T_H$ one would, in a spatially homogeneous way, 
need to invest an infinite amount of energy into the system 
which is impossible. By an (externally induced) violation of 
spatial homogeneity and thus by a sacrifice of thermal 
equilibrium the system may, however, condense densly packed (massless) 
vortex intersection points into a new ground state. The latter's excitations exhibit a 
power-like density of states and thus are described by a 
finite partition function. This is the celebrated Hagedorn 
transition approached from below \cite{Hagedorn1965}.

\section{Thermodynamical quantities throughout 
the deconfining and preconfining phase} 

\begin{figure}
\begin{center}
\leavevmode
\leavevmode
%\epsffile[80 25 534 344]{}
\vspace{4.5cm}
\includegraphics{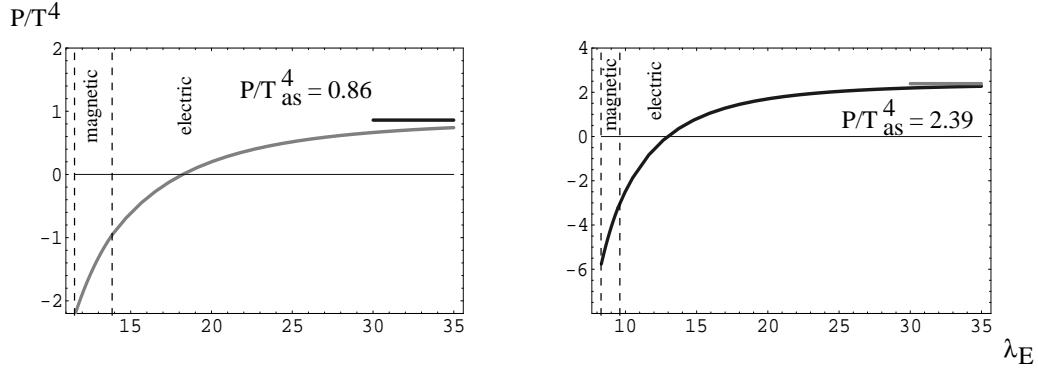}
\end{center}
\caption{\protect{$\frac{P}{T^4}$ as a function of temperature 
for SU(2) (left panel) and SU(3) (right panel). 
The horizontal lines indicate the respective asymptotic values, 
the dashed vertical lines are the phase boundaries.  
\label{pressure}}}      
\end{figure}

\begin{figure}
\begin{center}
\leavevmode
\leavevmode
%\epsffile[80 25 534 344]{}
\vspace{5.5cm}
\includegraphics{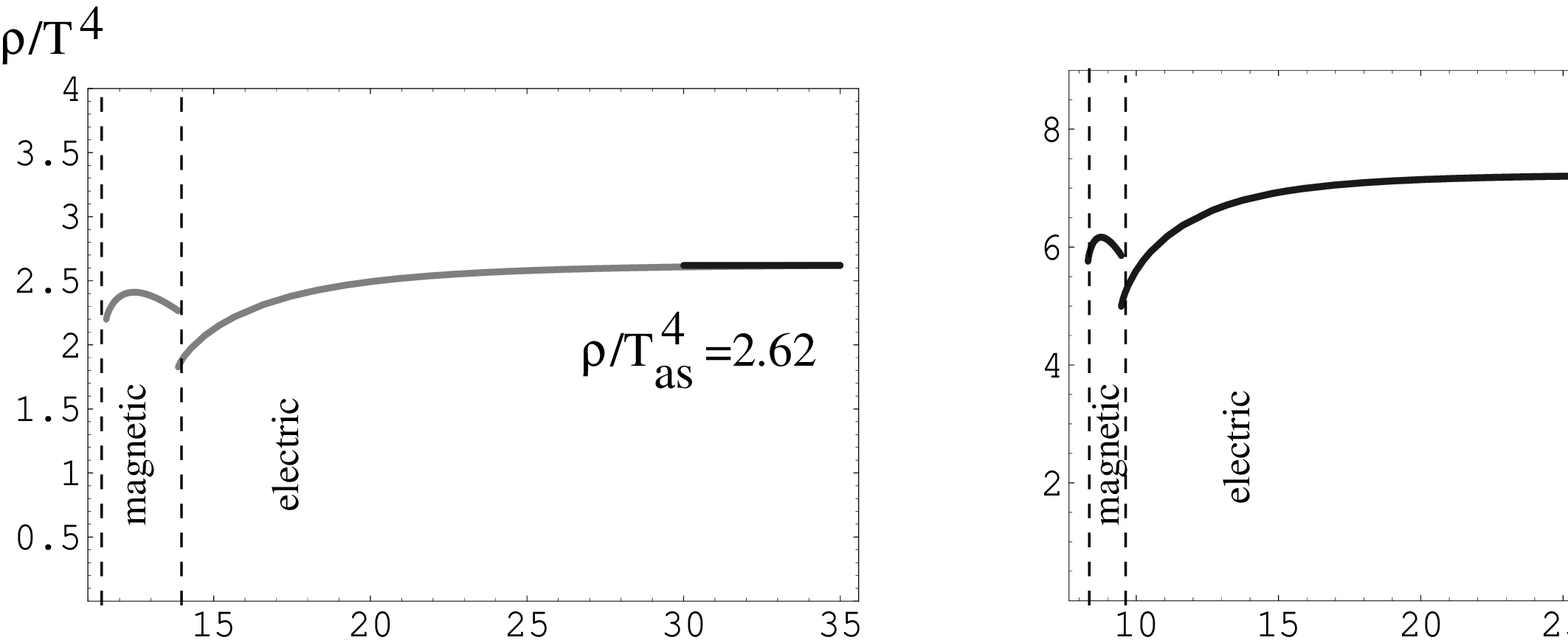}
\end{center}
\caption{$\frac{\rho}{T^4}$ as a function of temperature for SU(2) (left panel) and SU(3) (right panel). 
The horizontal lines indicate the respective asymptotic values, the dashed 
vertical lines are the phase boundaries.\label{rho}}      
\end{figure}

\begin{figure}
\begin{center}
\leavevmode
\leavevmode
%\epsffile[80 25 534 344]{}
\vspace{5.0cm}
\includegraphics{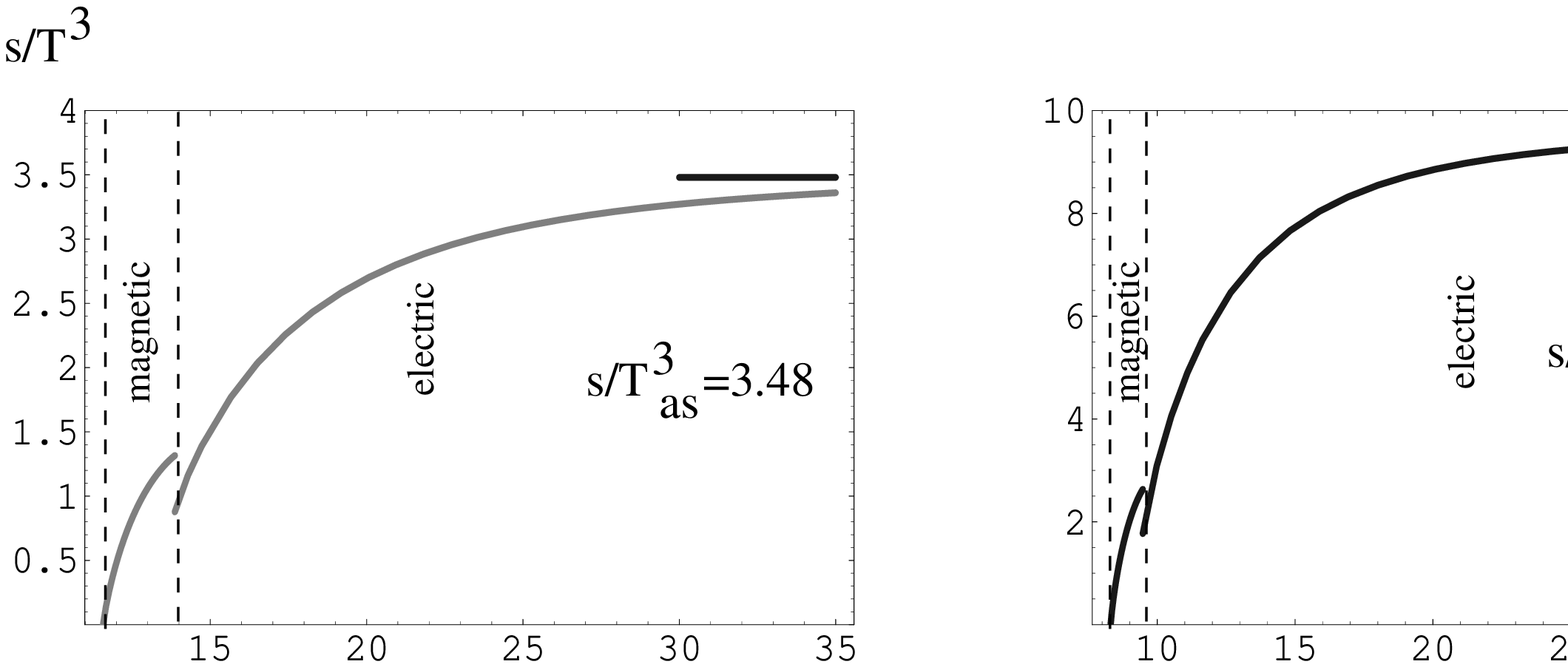}
\end{center}
\caption{$\frac{S}{T^3}$ as a function of temperature for SU(2) (left panel) and SU(3) (right panel). 
The horizontal lines signal the respective asymptotic values. \label{ST3}}      
\end{figure}

Here we indicate the temperature dependence of the pressure $P$ (Fig.\,\ref{pressure}), 
the energy density $\rho$ (Fig.\,\ref{rho}), and the entropy density $S$ (Fig.\,\ref{ST3}).
%***********************
%************************
%***********************
%************************
%***********************
%************************ 
Notice the negativity of the pressure at low temperatures. 
Notice also that all thermodynamical quantities reach their Stefan-Boltzmann 
limits very rapidly. At $T_{c,M}$ the pressure is discontinuous: It is positive and 
very large shortly below $T_{c,M}$ 
(Hagedorn transition). The entropy density vanishes at $T_{c,M}$: 
At this point the SU(2) or the SU(3) Yang-Mills theory only generates 
a cosmological constant.   

\section{Some implications: SU(2)$_{\tiny\mbox{CMB}}\stackrel{\tiny\mbox{today}}=$U(1)$_Y$} 

We only discuss implications arising from a modification of the gauge group 
describing photon propagation. Other implications are briefly discussed 
in \cite{Hofmann2005} and \cite{GiacosaHofmann2005}.

In \cite{Hofmann2005,SchwarzHofmannGiacosa2006,Hofmann20051,GiacosaHofmann2005} 
we have discussed why the U(1) gauge symmetry of electromagnetism is likely 
to have an SU(2) instead of an U(1)$_Y$ progenitor. 
To comply with the observational fact, that the photon is massless 
and unscreened \cite{Williams1971} within the present 
cosmological epoch, $T_{c,E}$ must coincide with 
the temperature $T_{\tiny\mbox{CMB}}\sim 2.7\,$K of the 
cosmic microwave background (CMB). Hence the 
name SU(2)$_{\tiny\mbox{CMB}}$. 

There emerges a number of unexpected 
results when subjecting the dynamics of SU(2)$_{\tiny\mbox{CMB}}$ to a 
thermodynamical treatment. 

First, one deduces 
that the present ground-state energy density of 
SU(2)$_{\tiny\mbox{CMB}}$ is less than 0.4\% of the measured 
value of today's density in dark energy \cite{GiacosaHofmann2005}. 
Thus there must be an additional mechanism providing for the latter. 
A serious possibility is that a Planck-scale axion field, coupling 
to the topological defects of SU(2)$_{\tiny\mbox{CMB}}$ and thereby acquiring a 
tiny mass ($m_{\tiny\mbox{axion}}\sim 10^{-36}\,$eV), 
is caught at the slope of its potential by cosmological 
friction. This axion field owes its existence to the chiral 
anomaly \cite{axialanomaly} taking place due to integrated-out 
chiral fermions as the temperature of the Universe fell below the 
Planck mass $M_P\sim 1.2\times 10^{19}\,$GeV. These fermions may have emerged because 
an SU(2) or an SU(3) gauge theory of Yang-Mills scale $\Lambda\sim M_P$ 
went confining. A (to-be-investigated) possibility is 
that all SU(2) or SU(3) gauge symmetries together with their Yang-Mills scales, 
describing the matter content of our (four-dimensional) Universe, 
were set by a dynamical symmetry breakdown: SU($\infty)\to$ gravity + matter. 
The Planck mass would then be associated with the 
Yang-Mills scale of SU($\infty$). The Planck-scale 
axion would trigger CP-violation in particle creation 
whenever an SU(2) or an SU(3) factor becomes nervous, that is, 
close to a Hagedorn transition 
(matter asymmetry \cite{Sakharov1967}).  

Second, one derives that due to the present cosmological expansion, mainly driven 
by the axion field, the photon remains massless only for a period $\Delta t_{m_\gamma=0}$ 
of at most 2\,billion years \cite{GiacosaHofmann2005}: After the 
time $\Delta t_{m_\gamma=0}$ has elapsed 
the photon acquires a Meissner mass (transition from deconfining to preconfining phase), 
and the ground-state of the Universe becomes superconducting. 
To obtain a tighter upper estimate for $\Delta t_{m_\gamma=0}$ 
one would have to relate the strength of intergalactic magnetic 
fields to the exact point in the phase diagram 
of SU(2)$_{\tiny\mbox{CMB}}$ corresponding to the 
present supercooled state of the Universe. 

\begin{figure}
\begin{center}
\leavevmode
\leavevmode
%\epsffile[80 25 534 344]{}
\vspace{5.8cm}
\includegraphics{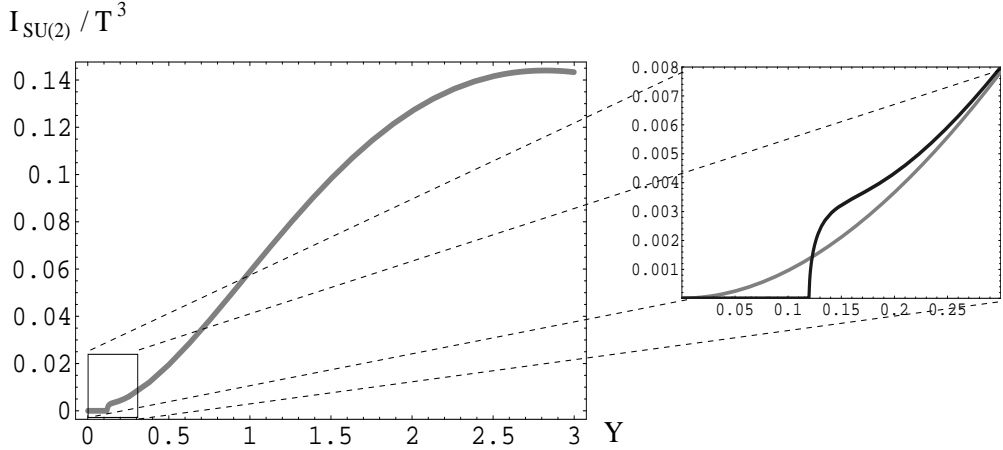}
\end{center}
\caption{\protect{\label{Fig-5}}Dimensionless black-body spectral power 
$\frac{I_{\tiny\mbox{SU(2)}}}{T^3}$ as a function of the dimensionless frequency 
$Y\equiv\frac{\omega}{T}$. The black curve in the magnified region depicts the modification of 
the spectrum as compared to $\frac{I_{\tiny\mbox{U(1)}}}{T^3}$ (grey curve) 
for $T=10\,$K.}      
\end{figure}

Third, if photon propagation is described by SU(2)$_{\tiny\mbox{CMB}}$ rather 
than U(1)$_Y$ then a visible modification of black-body spectra is 
predicted \cite{SchwarzHofmannGiacosa2006,Nature2006} 
for temperatures not much above $T_{\tiny\mbox{CMB}}$, 
see Fig,\,\ref{Fig-5}. 
%***********************
%************************ 
The spectral gap at low frequencies may explain why cold 
(brightness temperature $T_B=5\cdots 10\,$K) and dilute 
(particle distance $\sim 1\,$cm) clouds in between the spiral 
arms of the outer galaxy are composed of {\sl atomic} instead of molecular 
hydrogen and why such a situation is stable \cite{BruntKnee2001,Dickey2001,Dickey}. 
If SU(2)$_{\tiny\mbox{CMB}}$ is experimentally supported by the 
observation of the spectral gap in the black-body spectra at low temperatures 
then a contradition to the Standard Model's scenario for Big-Bang 
nucleosythesis arises. Namely, additional six relativistic 
degrees of freedom are present at the freeze-out temperature 
$T_{\tiny\mbox{freeze-out}}\sim 1\,$MeV for proton-to-neutron 
conversion implying a larger value of the Hubble parameter 
in comparison to the Standard Model. The proton-to-neutron ratio 
at freeze-out, which determines $T_{\tiny\mbox{freeze-out}}$, 
is rather tightly constrained by the relative abundance $Y_p$ 
of primordial $^4$He \cite{Eidelman2005}. One possibility 
to cure the mismatch in relativistic degrees of freedom 
would be to prescribe a larger value 
of the Fermi coupling $G_F$ at $T_{\tiny\mbox{freeze-out}}$. 
If the weak interactions and 
the emergence of the electron and its neutrino are 
described by an SU(2) gauge theory of Yang-Mills 
scale $\Lambda\sim 0.5\,$MeV then an enhancement of $G_F$ 
is, indeed, expected. Notice, however, that at 
$T_{\tiny\mbox{freeze-out}}\sim 1\,$MeV this theory 
is close to its Hagedorn transition; thus meaning 
that the synthesis of light elements would have 
taken place in a nonthermal environment.   

\section*{Acknowledgments}

I dedicate this paper to a number of very special women: my daughter Cattleya, 
my wife Karin, my mother Monika, my sister Katja, and my in-laws B\"arbel, 
Eva, and Nadja. The author would like to thank Francesco Giacosa and Markus Schwarz 
for useful comments on the manuscript.

\end{document}